# H4-Writer: A Text Entry Method Designed For Gaze Controlled Environment with Low KSPC and Spatial Footprint

**Raeid Saqur**
Dept. of Computer Science and Engineering
York University
Toronto, Ontario, Canada M3J 1P3
cs253208@cse.yorku.ca

**ABSTRACT**
This paper presents a new text entry technique, namely *H4-Writer,* designed for gaze controlled environments and aimed at reducing average KSPC and spatial footprint. It also presents an empirical evaluation of this proposed system by using three different input devices: *mouse*, *gamepad* and *eye tracker*. The experiment was conducted using 9 participants and the obtained data were used to compare the entry speeds, efficiency and KSPC of *H4-Writer* for all the devices. Over three blocks, the average entry speed was 3.54 wpm for the mouse, 3.33 wpm for the gamepad and only 2.11 wpm for the eye tracker. While the eye tracker fared poorly compared to the mouse and the gamepad on entry speed, it showed significant improvement in entry speed over progressing blocks indicating increase in entry speed with practice had a full longitudinal study was conducted . The average KSPC of all the three devices over all the text phrases entered was 2.62, which is significantly lower compared to other hand writing recognizing text entry techniques like *EdgeWrite*. An analysis of the blocks revealed improvement in error rate, efficiency and KSPC values with progressing block numbers as the participants got more acclimatized with the *key codes* for corresponding characters.

**ACM Classification Keywords**
H5.2 [Information interfaces and presentation]: User Interfaces – evaluation/methodology.

**INTRODUCTION**
Gaze gesture-based text entry techniques using eye trackers have evolved significantly over the years and with the increasing availability of eye trackers in size and price range, the search for more efficient text entry methods that use eye trackers constitutes a major research area in human computer interaction.

Text input with eye gaze is and most likely will never be a mainstream activity [4] due to the inherent limitations associated with eye-typing [1]. However, gaze controlled text inputting could be useful to people with severe motor disability. Also, the idea can be extended to mobile contexts where people with normal control over their limbs can find eye gaze interaction useful [4].

On screen keyboards, used for text inputting using eye gazing, which have one to one mapping of keys to characters (and commands) usually have large spatial footprint on the screen due to the large number of keys used. Other text entry systems include hand writing recognizing gaze gesture based text entry methods, which usually have low spatial footprint, however, have high *KSPC (Key Strokes per Character)* [7, 12].

The paper proposes an alternative text entry method, *H4-Writer*, which was designed keeping gaze controlled environments in mind. This method addresses the issues with *KSPC* and large spatial footprint associated with conventional text entry methods used for gaze controlled environments.

The proposed method uses only four directional keys and hence can be implemented with relatively low spatial footprint both on screen and using hardware. The method also uses minimum redundancy codes to map each character (and command) to the keys based on the character's relative frequency of usage. This ensures relatively lower *KSPC*.

This paper describes *H4-Writer*'s iterative development process and an evaluation of the proposed method using three different input devices.

**Related Work**

The gestural text entry system *EyeWrite* proposed by Wobbrock et al. [12] is based on *EdgeWrite*'s unistroke alphabet previously developed for enabling text entry on PDAs, joysticks, trackballs, and other devices [11] . The main advantage of *H4-Writer* over *EyeWrite* is that *H4-Writer* has significantly lower average KSPC (2.32) than EdgeWrite (3.35) (Figure 1). The KSPC calculation for these two text entry systems was done by using technique prescribed by MacKenzie, I. S. [7].

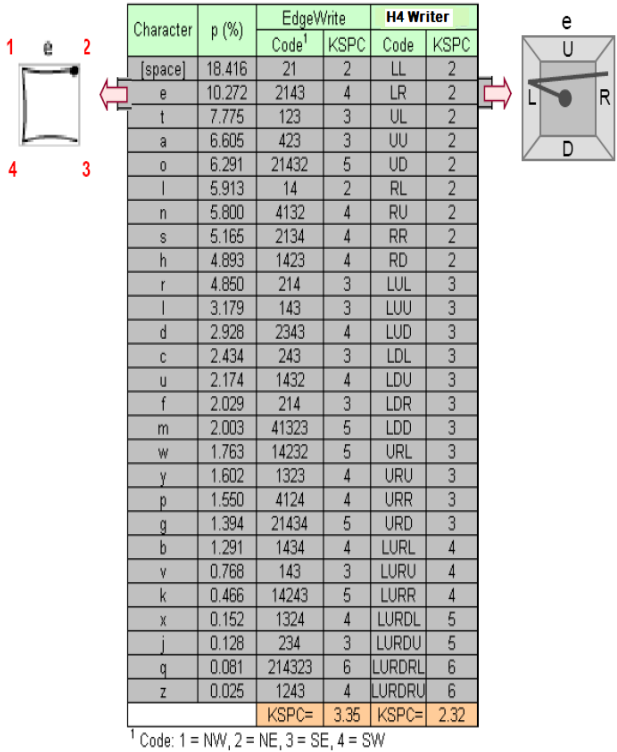

**Figure 1: KSPC of English alphabets using *EdgeWrite* and *H4-Writer*.**

Other gestural text entry techniques, such as *Graffiti* [3], impose problem of text input for people with motor impairments. Tremor and fatigue dramatically impact a user's ability to make smooth, accurate and controlled movements [5]. Due to tremor, it is difficult or impossible for some subjects to make character forms recognizable by a text entry system like *Graffiti*.

The proposed *H4-Writer* addresses this issue because it does not require the user to use letter-like gestures but simple dwell time based positioning and clicking gestures. This is described in more detail in the following section where the basic implementation of the H4-Writer has been discussed.

**H4-Writer**

The name of the proposed gaze-based text entry approach is derived from the fact that the system uses *Huffman coding* and four input keys. As briefly mentioned earlier, one main advantage of *H4-Writer* over other similar techniques is that it uses lower KSPC. This is achieved by deriving unique prefix, minimum redundancy codes for each letter in the English alphabet by using *Huffman Coding*.

The following figure (Figure 2) shows the relative frequencies of letters in the English language [6].

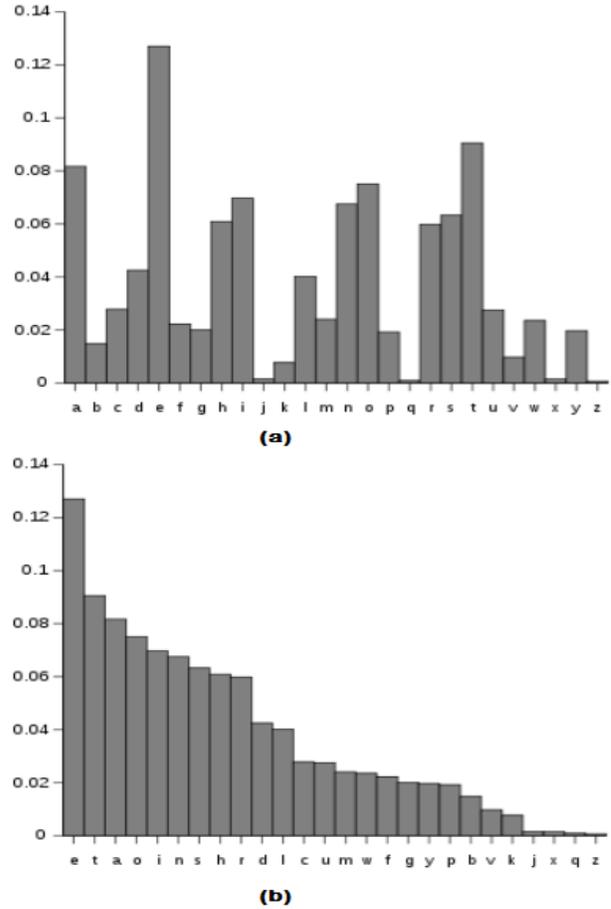

**Figure 2: (a) Relative letter frequencies (b) frequencies in descending order.**

Each alphabet is mapped to its code using these relative frequencies. So, the alphabets with higher relative frequencies will have shorter code and vice versa. For example, the following is a portion of the codes defined for English alphabets and some commands (Figure 3):

```
Code translation: L = Left; R = Right; U = Up; D = Down
            [Space]         LL
            [Bksp]          DD
            [Enter]         DR
            .......         ...
            e               LR
            t               UL
            .......         ...
            r               LUL
            .......         ...
            g               URD
            b               LURL
            v               LURU
            .......         ....
            q               LURDRL
            z               LURDRU
```

**Figure 3: Input key codes corresponding to letters and commands.**

Besides the alphabets, codes were derived for 'space', symbols and various commands. However, in this paper, we only consider the lower case English alphabets, space and other pertinent commands needed for general text entry for the experiment excluding symbols. As we can see from Figure 3, the letters with high relative frequency (for e.g. 'e') has shorter codes ('LL') compared to letters with low relative frequency (for e.g. 'q' = 'LURDRU').

An on screen GUI applet containing the four input keys was developed for evaluating this proposed system. This will be discussed in more detail in the apparatus section of this paper.

This paper mainly discusses the pilot research study aimed at evaluating the performance of the proposed H4-Writer using different input devices and the skill acquisition trend among novice users.

## METHOD

### Participants
Nine participants were recruited from the local university campus (8 male, 1 female; all of them right handed; mean age = 23.6 years). All participants were highly experienced in computer use (average computer usage = 8.6 hours/day), but none had previously used an eye tracker. All participants possessed a university degree or were currently enrolled in university. All subjects rated themselves as having either good or expert typing skills. All participants were familiar with using gamepads, however, none of them had used gamepads for text entry previously. Two participants reported some experience with *Graffiti*, a unistroke handwriting method used by Palm PDAs [3]. Due to their unfamiliarity with *H4-Writer*, all the participants were classified as '*novice*' users.

### Apparatus
The experiment was conducted in a quiet lab. Dim lighting was used while conducting the experiment using eye tracker and normal room lighting was used for the rest of the experiment.

*Input Devices*
This study relied on three different input devices. For text entry using eye gazing, a non-invasive *EyeTech TM3* eye tracker with a sampling rate of 50 Hz was used (Figure 4). The resolution was set to 1026 × 768 pixels (14.1 inch LCD screen).

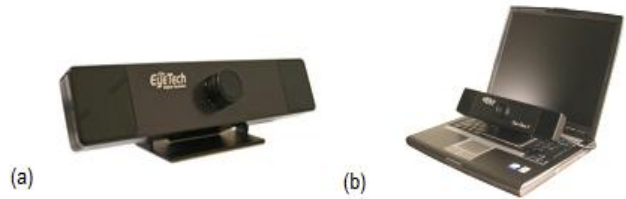

**Figure 4: (a) EyeTech TM3 (b) Laptop setup.**

For text entry using gamepad, a *Logitech* PC gamepad was used (Figure 5). The four right-hand side numbered buttons were mapped as the four directional keys used by *H4-Writer*.

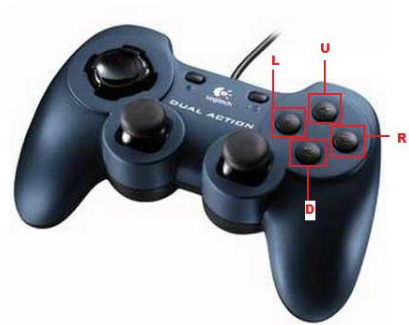

**Figure 5: Logitech PC Gamepad.**

For text entry using mouse, a standard size *Microsoft* wireless mouse was used (Figure 6).

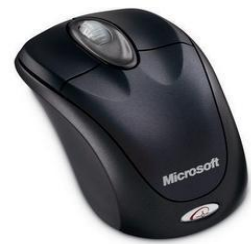

**Figure 6: Microsoft wireless mouse.**

*Application*

Figure 7 depicts the *H4-Writer* application used for the experiment. The topmost text area displayed the presented phrase, while the lower one displayed the transcribed text. The middle section contained the four directional key buttons, and all the alphabets and commands arranged in boxes adjacent to the directional keys according to their corresponding Huffman key codes. The bottom panel was used to display the results (dependent variables) after each entry was completed by the user.

The application was designed with the same "look and feel". This included audio feedback ("click") on a key press event, distinct audio feedback ("beep") if the wrong character was entered and a highlight effect on a rollover event when a key receives focus.

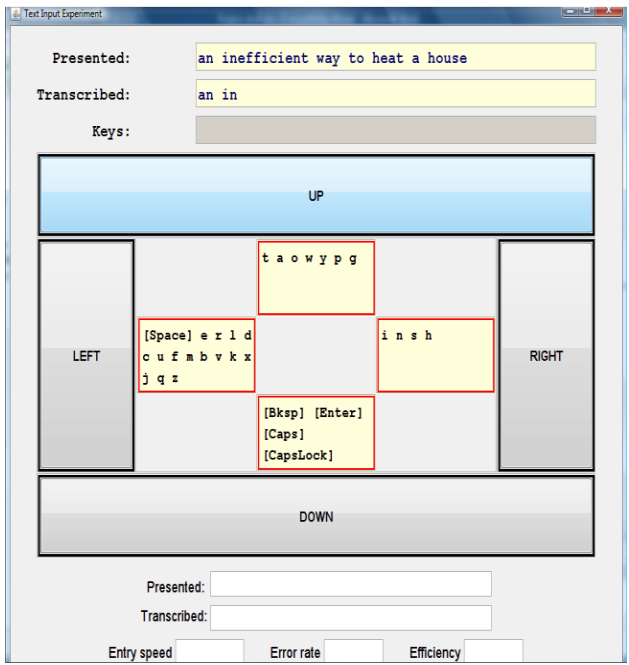

**Figure 7:** The *H4-Writer* application used for the experiment.

For example, to transcribe the letter '*e*', the corresponding code '*LR*' (Figure 3) needs to be keyed. The user does not need to remember this code. The first key to press is determined by searching for '*e*' in one of the four boxes containing the letters, then the key (button) adjacent to the box is clicked. The first button in this case is 'LEFT' (code: 'L'). This step is repeated until the letter (or command) to transcribe is the only letter left in the boxes. For the case of '*e*', the second and the last key to press is 'RIGHT' (code: 'R'). As soon as the last key is pressed, the letter is entered in the transcribed text area along with a distinct audio feedback. Figure 8 depicts this example scenario pictorially.

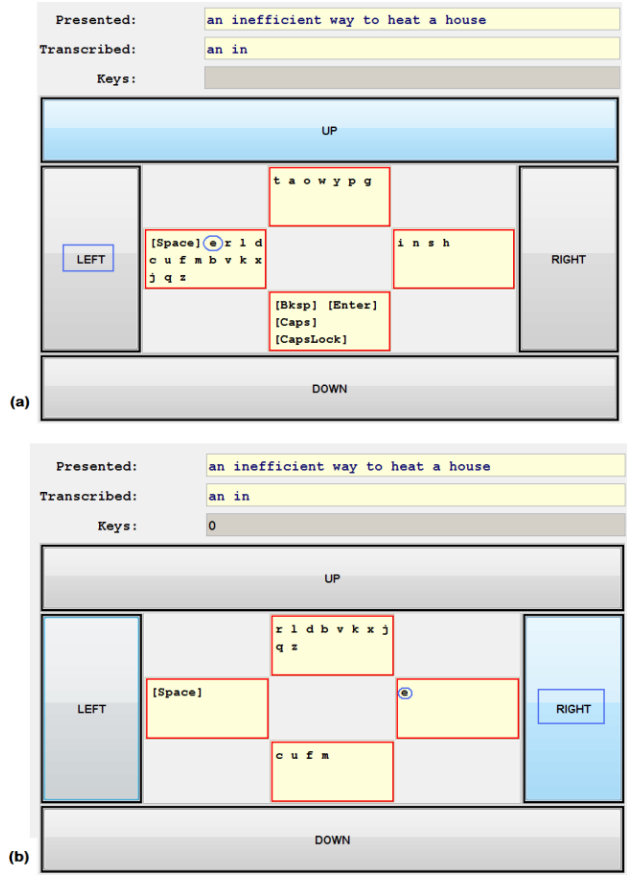

**Figure 8:** Entering letter '*e*' (a) first key 'L' (b) second key 'R'.

**Procedure**

Participants first signed an informed consent waiver then completed a short demographic questionnaire. Next, the participants were given basic instructions on how to enter text with each device. Before the participants did the experiment using the eye tracker, the eye tracker was calibrated to each participant using twelve calibration points. Participants were then given five to ten minutes of unguided practice time to become comfortable typing with the eye tracker using their assigned condition. For the experiment, they were instructed to proceed "as quickly and accurately as possible".

Participants performed the experiment while sitting down. They used their dominant hands for the mouse and the gamepad. For the eye tracker, the distance between the user and the screen was about 60 cm. Each participant entered three blocks of phrases for each input device; each block contained three different randomly chosen phrases from a 500-phrase set [9]. Each study lasted approximately 50 minutes.

Figure 9 depicts the basic experimental setup (a) and the setup for all the three input devices.

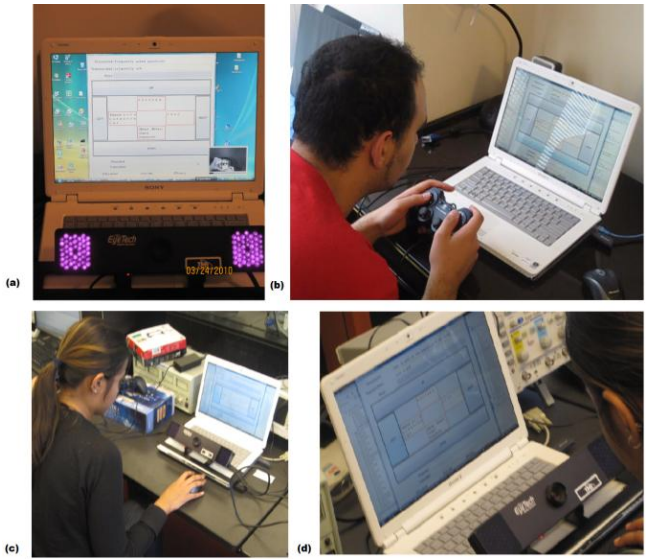

Figure 9: Experiment setup (a) basic setup (b) gamepad (c) mouse (d) eye tracker.

**Design**

The experiment was a 3 × 3 within-subjects design. There were two independent variables: input device (*mouse, gamepad, eye tracker*) and block (1-3). Each block contained three phrases of text entry. The dependent variables were entry speed (wpm), efficiency (%) and KSPC. Participants were divided into three groups with the order of input devices counterbalanced to neutralize learning effects [8]. In total, the number of phrases entered was 9 participants × 3 input techniques × 3 blocks × 3 phrases/block = 243.

## RESULTS AND DISCUSSION

**Performance**

Figure 10 shows the results for entry speed. On average, participants entered text at 3.54 wpm (SD = 0.79) using the mouse, 3.33 wpm (SD = 1.09) using the gamepad and 2.11 wpm (SD = 0.33) using the eye tracker. In relative terms, the *mouse* was 67.5% and the *gamepad* was 57.7% faster than the *eye tracker*, which indicates considerable difference.

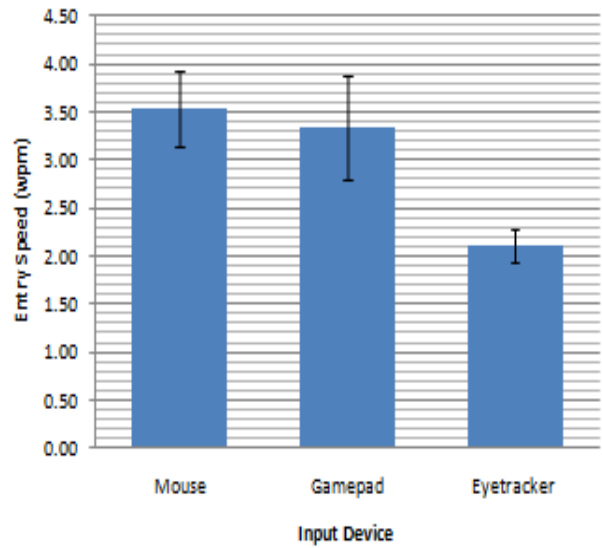

Figure 10: Entry speed vs. input device.

The effect on input device on entry speed was statistically significant ($F_{2,12} = 33.2$, $p < .0001$).

The entry speed results for the eye tracker (2.11 wpm) was much lower than the eye-typing speeds achieved with *EyeWrite* (5 wpm on average with novice users) [12], the gaze gesture based text entry system discussed in the '*Related Work*' section of this paper. Some probable reasons which attributed to this could be that in the longitudinal study for *EyeWrite*, the participants were allowed much longer acclimatization period with the system, and all the apparatus used for the study were fully optimized unlike this pilot study of *H4-Writer*. However, the learning curve for *H4-Writer* showed significant improvement rate of entry speed over longer use. This is discussed in more detail in a later part of this paper.

The results obtained suggest that using a mouse or a gamepad for *H4-Writer* yields significantly higher entry speed. The reasons for the significant difference is the directness of interaction using these two devices, and the interaction limitations associated with the use of eye trackers and gaze gesture based text entry [1].

Small improvements in the entry speed across the blocks were observed with all three of the input devices (Figure 11).

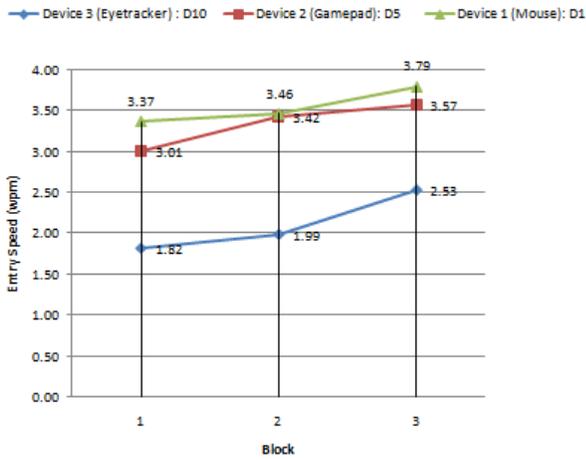

**Figure 11: Entry speed vs. block.**

The effect of block on entry speed was statistically significant ($F_{2,24} = 33.2$, $p < .0001$), even though there was a relatively small amount of text entered. Had a full longitudinal study been conducted, there likely would be more improvement with practice. Figure 12 depicts the learning curve over blocks. A trend line was added, based on the empirical data, projecting the improvement in speed (wpm) over 5 addition blocks. They indicate a strong improvement in speed over progressing blocks using all the three devices; however, more so using the *gamepad* and the *eye tracker* than the *mouse*. Note the higher $R^2$ values of the *gamepad* ($R^2 = 0.9812$) and the *eye tracker* ($R^2 = 0.8411$) compared to the mouse ($R^2 = 0.8145$). This is expected, because the *key codes* corresponding to the alphabets are learned and the participants get more acclimatized with the entry system over progressing blocks.

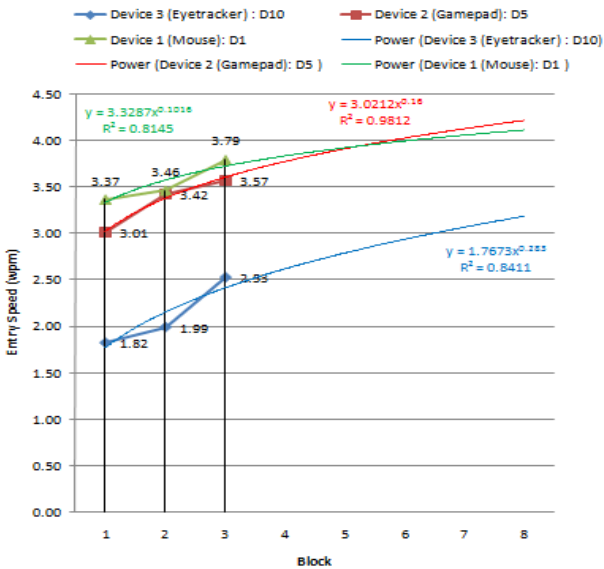

**Figure 12: Text entry speeds with each input device over 3 blocks and added trend line projecting 5 more blocks.**

In the following sections, we will discuss the other dependent variables of the experiment.

### Error Rate
*Uncorrected Errors*

In this experiment, the participants were allowed to correct any mistakes made while entering the texts if they wanted to do so. Thus it was possible to make a significant amount of errors while entering a phrase and yet have the number of (uncorrected) errors as 0. Hence, although empirical data for uncorrected errors were collected and recorded, they do not have statistical significance. For example, the effect of input device on uncorrected error rate was not statistically significant ($F_{2,12} = 1.11$, $p > .05$). Also, the effect of block on uncorrected error rate was not statistically significant ($F_{2,24} = 0.57$, $p > .05$).

The *corrected errors* were taken into account while calculating the *efficiency*. The following section discusses *efficiency* in more detail.

### Efficiency

Efficiency was calculated based on the number of corrected errors. The *efficiency* decreased with increased number of *corrected errors*.

The effect of input device on efficiency was not statistically significant ($F_{2,24} = 2.96$, $p > .05$). However, the effect of block on efficiency was somewhat statistically significant ($F_{2,24} = 6.01$, $p < .05$). This is expected because the number of corrected errors decreased with progressing block numbers as the participants got more acclimatized with the *key codes* corresponding to the letters and the commands. Figure 13 depicts the increase in efficiency with progressing blocks.

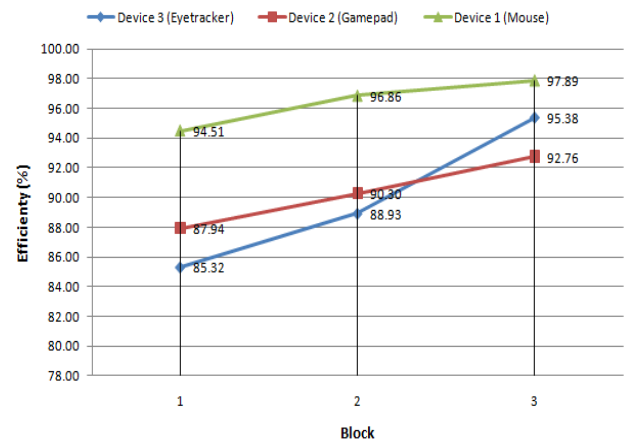

**Figure 13: Efficiency vs. block.**

Figure 14 depicts the projected trend in efficiency over five extra blocks. These show improvement in efficiency over progressing blocks with all three input devices. Had a full longitudinal study been conducted, there likely would be more improvement in efficiency with practice.

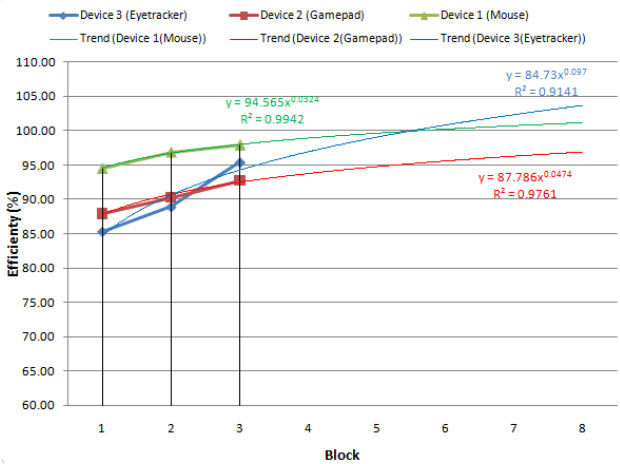

**Figure 14: Efficiency with each input device over 3 blocks and added trend line projecting 5 more blocks.**

### KSPC

The average overall KSPC of all the phrases entered in the experiment using all the three input devices was 2.63, which is still significantly lower compared to average KSPC of *EdgeWrite* [11].

The effect of input device on KSPC was not statistically significant ($F_{2,12} = 2.39$, $p > .05$). This is expected because KSPC is mainly dependent on the *key codes* used for all the alphabets. However, the effect of block on KSPC was somewhat statistically significant ($F_{2,24} = 5.04$, $p < .05$). Figure 15 shows the average KSPC for each of the input devices with increasing block numbers. This is expected because the number of (corrected) errors decreases and efficiency increases with more practice. Since KSPC is dependent on the key size, and key size is dependent on the efficiency and error rate, hence, KSPC decreases slightly with progressing blocks.

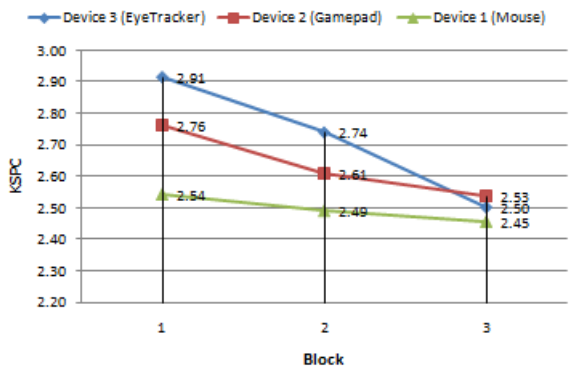

**Figure 15: KSPC vs. block.**

### CONCLUSION

In general, using *H4-Writer* with mouse and gamepad yielded better performance and KSPC using just three blocks. However, eye tracker showed significant improvement both in terms of performance and KSPC with progressing blocks as the participants got more acclimatized with the *key codes* used in the proposed system.

This work is important since the new proposed text entry method is novel and addresses two key issues (KSPC and spatial footprint) in text entry systems. Future work is intended to conduct a full longitudinal study of *H4-Writer* using fully optimized apparatus.

### ACKNOWLEDGEMENT
The author thanks all the participants who volunteered to take part in this experiment. The author also thanks Professor I.S. Mackenzie for advising and overseeing all aspects of this experiment.